**Title: Magnetic Field-Enhanced Graphene Superconductivity with Record Pauli-Limit Violation**

**Authors:**
Jixiang Yang[1*], Omid Sharifi Sedeh[2*], Chiho Yoon[3*], Shenyong Ye[1*], Henok Weldeyesus[2], Armel Cotten[2], Tonghang Han[1], Zhengguang Lu[1], Zach Hadjri[1], Junseok Seo[1], Lihan Shi[1], Emily Aitken[1], Prayoga P Liong[1], Zhenghan Wu[1], Mingchi Xu[2], Christian Scheller[2], Mingyang Zheng[4], Rasul Gazizulin[4,5], Kenji Watanabe[6], Takashi Taniguchi[7], Dominique Laroche[4], Mingda Li[8], Fan Zhang[3†], Dominik M. Zumbühl[2†], Long Ju[1†]

**Affiliations:**
1. Department of Physics, Massachusetts Institute of Technology, Cambridge, MA, USA
2. Department of Physics, University of Basel, Basel, Switzerland
3. Department of Physics, The University of Texas at Dallas, Richardson, TX, USA
4. Department of Physics, University of Florida, Gainesville, FL, USA
5. National High Magnetic Field Laboratory High B/T Facility, University of Florida, Gainesville, FL, USA
6. Research Center for Electronic and Optical Materials, National Institute for Materials Science, Tsukuba, Japan
7. Research Center for Materials Nanoarchitectonics, National Institute for Materials Science, Tsukuba, Japan
8. Department of Nuclear Science and Engineering, Massachusetts Institute of Technology, Cambridge, MA 02139, USA.

**Abstract:**
Spin-polarized superconductors offer a rare platform for studying electronic correlations, but few candidate systems have been experimentally confirmed to date. Here, we report the observation of a spin-polarized superconducting state, denoted SC5, in $WSe_2$-proximitized rhombohedral trilayer graphene. At in-plane magnetic field $B_{||}$ = 0 T, SC5 has a critical temperature of 68 mK and an out-of-plane critical magnetic field of only 12 mT. Surprisingly, these values are significantly enhanced as $B_{||}$ increases, and the superconductivity persists to $B_{||}$ = 8.8 T. This value corresponds to a record-high Pauli-limit violation ratio of at least 80 among all superconductors, while the true critical field is beyond the limit of our instrument. We conclude that SC5 experiences a canting crossover from Ising-type to spin-polarized superconductor with increased $B_{||}$.

**Main Text:**
As explained by the extraordinarily successful BCS theory(*1*), most of the existing superconductors are formed by the condensation of spin-singlet Cooper pairs. Spin-triplet or spin-polarized superconductivity, although of great theoretical interests, remains experimentally elusive due to its lower robustness against disorder compared to conventional BCS superconductivity(*2*, *3*). The hallmark of spin-polarized superconductivity is its extreme robustness against magnetic field that extends well beyond the Pauli-limit, especially in two-dimensional superconductors where the in-plane magnetic field renders a negligible orbital effect. Signatures of spin-polarized superconductivity have been observed in several candidate systems, e.g., perovskite $Sr_2RuO_4$(*4–6*), heavy-fermion superconductors(*7–13*), and Bernal bilayer graphene(*14*). However, the nature of these superconductors has been under debate, with alternative explanations suggested by new experiments(*15–19*). Therefore, the pursuit of spin-polarized superconductors with unambiguous experimental signatures remains critical, driven not only by fundamental scientific interest but also by the great potential for dissipationless spintronic and quantum device applications(*20*, *21*).

In recent years, rhombohedral-stacked graphene (RG) has emerged as a new platform for studying superconductivity(*22–28*), among other novel correlated and topological electron phenomena(*29–42*). More than ten different superconducting states have been reported in RG with various layer numbers and device configurations, providing an unprecedented opportunity to study the interplay between superconductivity and fractional electron physics, electronic crystals, spin-orbit couplings, moiré effects, etc., in a simple crystalline carbon material system. In particular, stacking RG on transition metal dichalcogenides (TMD) has been shown to effectively induce proximity spin-orbit couplings (SOC)(*24*, *26*, *43–48*) and modulate spontaneously symmetry-broken ground states. This extra experimental knob opens up a new territory for the study of superconductivity. For example, the Ising SOC breaks the inversion symmetry and thus allows the mixture of spin-singlet and spin-triplet superconductivity in principle(*49–51*), and the induced spin-valley-unpolarized half-metal state could possibly host Ising-type superconductivity(*52–54*) that is conceivably more robust to magnetic fields than conventional BCS superconductivity.

In this work, we report electronic transport studies of rhombohedral trilayer graphene (RTG) encapsulated by TMDs. We observed a new superconducting state, denoted SC5, near the quarter-metal phase on the hole-doping side. Strikingly, by applying an in-plane magnetic field up to 8.835 T, SC5 shows no sign of being suppressed and instead becomes progressively enhanced. This corresponds to a record-high Pauli-limit violation ratio (PVR) of 80, which is solely limited by the maximal field of the magnet and is likely to be much higher. We further study the parent states of SC5 via fermiology analysis and discuss the evolution of its spin configuration. We conclude that the in-plane magnetic field facilitates spin-polarized pairing by introducing parallel components of spins, and SC5 is enhanced through spin canting from its Ising-type nature at zero magnetic field due to the dominance of spin-triplet superconducting pairing. The data shown in the main text are from device D1, while the data from a second device D2 and a control device D3 are included in Fig. S1.

**A new superconducting state SC5**

The RTG devices were fabricated by encapsulating the active region between bilayer $WSe_2$ flakes (see the right inset in Fig. 1A), with graphite top and back gates that independently control the carrier density $n$ and displacement field $D$. Figure 1A shows the longitudinal resistance $R_{xx}$ (defined as the four-terminal differential resistance $\mathrm{d}V_{xx}/\mathrm{d}I$ at zero DC current) as a function of the carrier density $n$ and vertical displacement field $D$ at base electron temperature of ~20 mK. As we have shown previously, the $WSe_2$ layers induce spin-orbit coupling in RTG via the proximity effect and impact superconducting states SC1, SC3, and SC4 differently(*24*). Here, improved electronic filtering and noise suppression reveal a previously unobserved zero-resistance state, indicated by the red arrows. Figure 1B shows a zoomed-in plot near this new zero-resistance state, which corresponds to the red box in Fig. 1A. Figure 1C presents $\mathrm{d}V_{xx}/\mathrm{d}I$ as a function of the direct current $I_{DC}$, measured at the orange dot position in Fig. 1B. The differential resistance remains at zero for small DC currents, then exhibits two strong peaks around threshold current $I_{DC}$ = ± 10 nA, which we define as the critical current $I_c$.

Such non-linear *I-V* behavior confirms that the new zero-resistance state is a superconductor. We refer to it as SC5, following the naming convention in previous RTG papers(*22–24*). SC5 resides near the tip of the quarter-metal (QM) phase(*24*, *26*, *39*, *55*) and is split into two regions that are barely connected. Since our devices are designed to maintain inversion symmetry, the phase diagrams at positive (Fig. 1B) and negative (Fig. S2G) displacement fields are almost identical, except that SC5 is more well-developed in the positive-*D* side, which could originate from a tiny asymmetry unintentionally generated during sample fabrication. Therefore, we will focus on the positive-*D* side in the remainder of this paper. The SC5 state is also observed in device D2 with a similar sample structure, as shown in Fig. S1B. We also note that in the control device D3, where RTG is only proximitized by a monolayer $WSe_2$ flake on the top side, SC5 is not observed, regardless of whether the electrons are polarized toward or away from the graphene/TMD interface, as shown in Fig. S1E & F.

Figure 1D shows the dependence of $dV_{xx}/dI$ on the direct current $I_{DC}$ as a function of out-of-plane magnetic field $B_\perp$, which again confirms superconductivity and determines that its out-of-plane critical field $B_{c\perp}$ is around 12 mT. Figure 1E displays the temperature dependence of the $R_{xx}$ in SC5 state (same as Fig. 1C & D, measured at the orange dot position in Fig.1B), which exhibits an abrupt drop around $T_c$ = 68 mK. The Berezinskii–Kosterlitz–Thouless transition temperature(*56*) $T_{BKT}$ = 59 mK can be extracted from fitting $V_{xx} \sim I^3$ from the inset of Fig. 1E. The critical temperature $T_c$, the critical current $I_c$, and the out-of-plane critical magnetic field $B_{c\perp}$ of SC5 are all similar to the typical values of other superconductors in crystalline graphene-based systems. However, as we will show below, SC5 behaves strikingly differently from all other two-dimensional superconductors under an in-plane magnetic field.

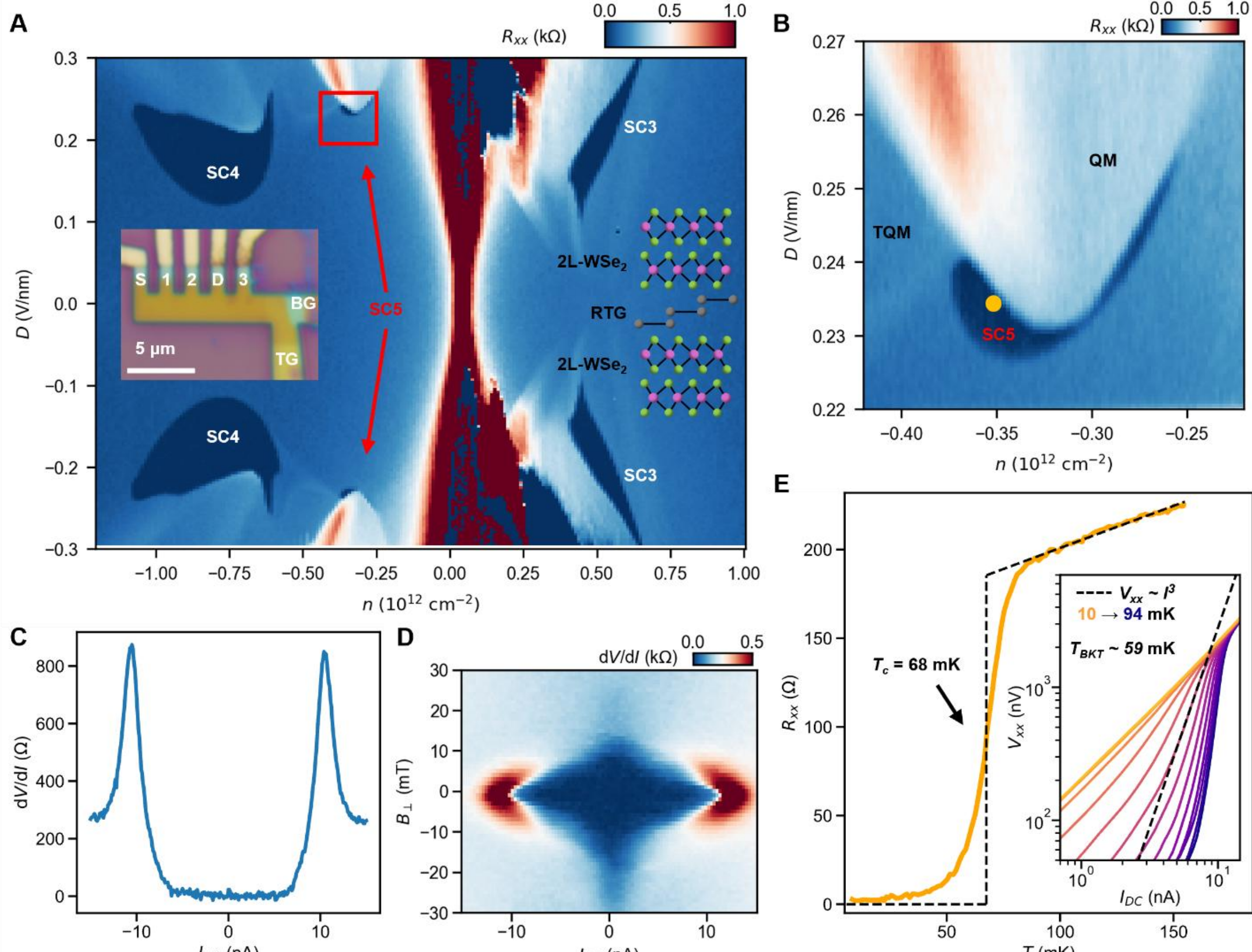

***Fig. 1: Superconducting State SC5.***
***(A)*** *$R_{xx}$ as a function of charge density n and gate displacement field D. In addition to previously reported SC3 & SC4, a new superconducting state SC5 is discovered and highlighted by the red arrows. Left inset: optical image of the device, where TG and BG correspond to the top and bottom gates, respectively. Right inset: illustration of the device structure. RTG is encapsulated by two flakes of bilayer 2H-$WSe_2$ that are aligned parallel with each other to preserve overall inversion symmetry.* ***(B)*** *Zoomed-in n-D plot near SC5, corresponding to the red square region in (A). SC5 resides near the tip of a quarter-metal (QM) phase and neighbors a three-quarter-metal (TQM) phase.* ***(C)*** *Differential resistance $dV_{xx}/dI$ as a function of $I_{DC}$, measured at n & D corresponding to the orange dot in (B).* ***(D)*** *Differential four-terminal resistance $dV_{xx}/dI$ as a function of the direct current $I_{DC}$ and out-of-plane magnetic field $B_\perp$, measured at the same position as in (C). The out-of-plane critical magnetic field $B_{c\perp}$ is around 12 mT.* ***(E)*** *Temperature dependence of $R_{xx}$ measured at the same position as in (C). The dashed line shows where $R_{xx}$ falls to half of the normal state resistance $R_{xx,n}(T)$, which defines the transition temperature $T_c$ =68 mK. Inset: $V_{xx}$ as a function of direct current $I_{DC}$ (for the same state) as a function of temperature. The Berezinskii-Kosterlitz-Thouless transition temperature $T_{BKT}$ is ~ 59 mK, as shown in the inset.*

## Enhancement of SC5 under in-plane magnetic field

Figure 2, A – C, shows the longitudinal resistance as a function of *n* and *D* in the same region of Fig. 1B, measured at in-plane magnetic fields $B_{\parallel}$ = 1, 4, and 7 T, respectively. The phase diagram of SC5 at $B_{\parallel}$ = 1 T is almost identical to that at zero magnetic field. Figure 2B, surprisingly, shows that the two weakly linked regions of SC5 merge into one, and this merged region further enlarges in Fig. 2C. Such an enlargement of the superconducting phase space is against the general lore of the BCS theory that magnetic field would suppress superconductivity. Figure 2D summarizes the evolution of SC5 from $B_{\parallel}$ = 0 to 7 T, whose boundary is defined as the contour of $R_{xx}$ = 100 Ω, approximately half of the normal state resistance. The boundary only slightly changes from $B_{\parallel}$ = 0 to 3 T, and starts to expand drastically from 4 to 7 T. Figure 2E and F show the continuous evolution of a constant-$n$ and a constant-$D$ linecut with step size $\Delta B_{\parallel}$ = 0.1 T. In both plots, the boundary between SC5 and the QM phase remains nearly unchanged for $B_{\parallel} \leq 3$ T; above that value, SC5 expands toward the QM phase.

In addition to the obvious enlargement of its $n$-$D$ range, we also investigated other characteristics of superconductivity under applied $B_{\parallel}$. Figure 2G shows critical temperatures $T_c$ at several different in-plane magnetic fields. $T_c$ remains at around 70 mK for $B_{\parallel} \leq 3$ T but increases to around 95 mK for $B_{\parallel}$ = 5 T and 7 T. Figure 2H shows the differential resistance as a function of direct current at $B_{\perp}$ = 0 T and various $B_{\parallel}$. As $B_{\parallel}$ increases, the resistance peaks shift to a higher critical current $I_c$. Figure 2I shows similar measurements to those in Fig. 2H but at $B_{\perp}$ = 15 mT. At $B_{\parallel} \leq 3$ T, the 15 mT out-of-plane magnetic field can eliminate the resistance peaks and suppress SC5, while the resistance peaks at $B_{\parallel}$ = 5 & 7 T can still be observed under the same 15 mT out-of-plane magnetic field. We further determine the out-of-plane critical magnetic field increases to ~ 25 mT at $B_{\parallel}$ = 7 T (see Fig. S3D), contrasting the critical value of 12 mT at $B_{\parallel}$ = 0 T.

Combining the enlarged $n$-$D$ range, the increased critical temperature, critical current, and critical out-of-plane magnetic field, we conclude that SC5 is enhanced by the applied in-plane magnetic field up to 7 T. In a second round of measurements using a different pair of contacts in the same sample (see Supplementary Materials for details), as shown in Fig. S6B, we were able to achieve an in-plane magnetic field of 8.835 T, where SC5 still persists. Such robustness of superconductivity against in-plane magnetic field is highly unusual: according to the BCS theory(*57–59*), the superconductivity is expected to be fully suppressed at the Pauli limit of $T_c \times$ 1.86 T/K ≈ 126 mT for SC5, which is two orders of magnitude smaller than we have observed. We stress that 8.835 T is only the achievable limit of the magnet we used, while the true critical in-plane magnetic field might be even higher, corresponding to an even higher PVR. Furthermore, the enhancement of SC5 by in-plane magnetic fields is in sharp contrast to all other two-dimensional superconductivities that have been reported so far.

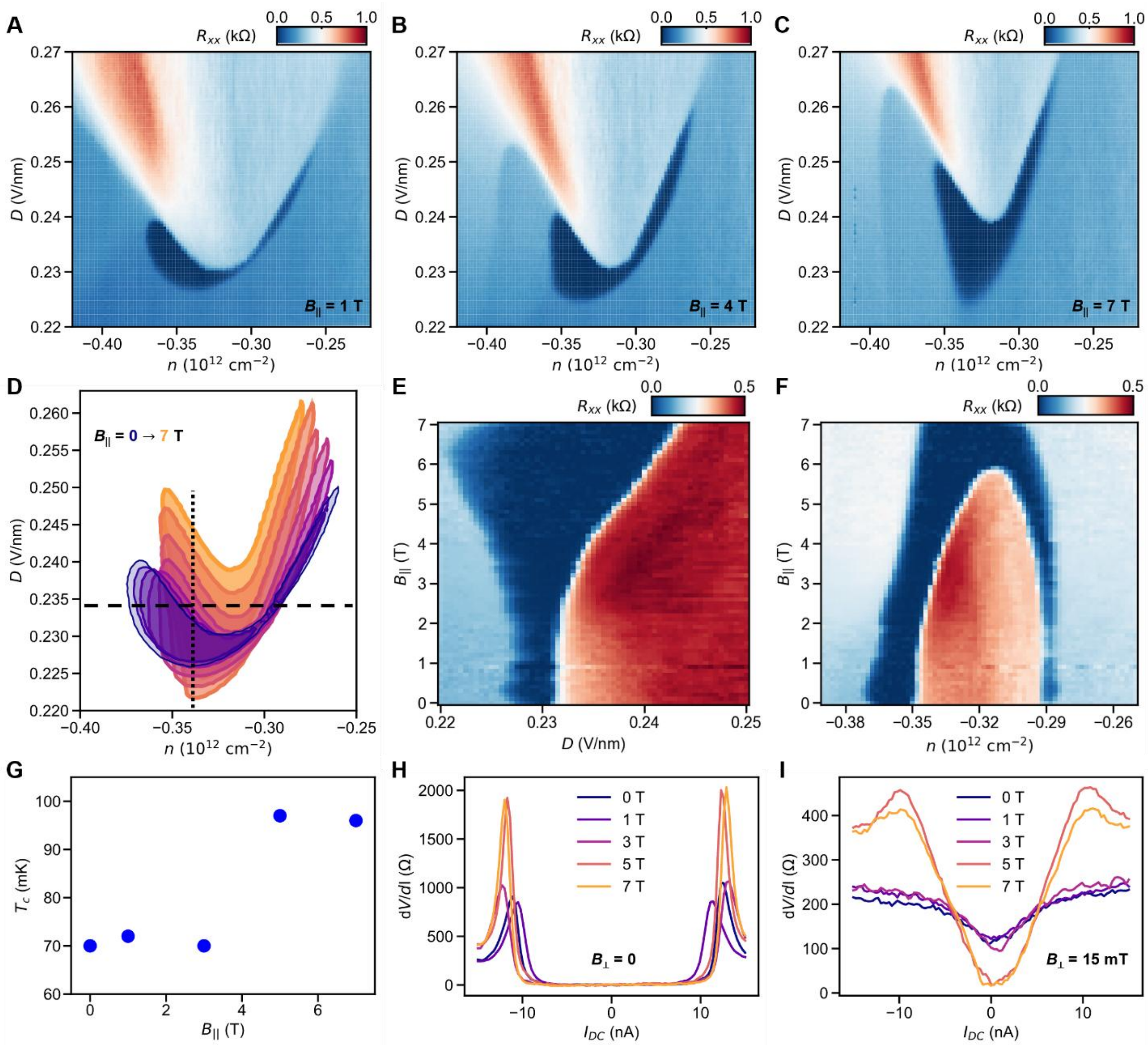

***Fig. 2: Enhancement of SC5 under in-plane magnetic field.*** *(**A** – **C**) $R_{xx}$ as a function of n and D measured in the same region as in Fig. 1b at in-plane magnetic field $B_{||}$ = 1, 4, and 7 T, respectively. As $B_{||}$ increases, SC5 expands in area in the n-D phase diagram. (**D**) The boundary of SC5 in the n-D phase diagram at $B_{||}$ = 0 to 7 T, which is defined as the contour of $R_{xx} = R_{xx,n}/2 \approx 100\ \Omega$. (**E** and **F**) The dependence of $R_{xx}$ on $B_{||}$ up to 7 T, along the dotted (E) and the dashed (F) black lines in (D). SC5 evolves continuously from zero magnetic field to $B_{||}$ = 7 T and expands against the quarter-metal (QM) state starting from 3 T. (**G**) The critical temperature $T_c$, defined as the highest $T_c$, measured at $B_{||}$ = 0, 1, 3, 5, and 7 T. $T_c$ remains around 70 mK for $B_{||} \leq 3$ T, but rises to around 95 mK for $B_{||}$ = 5 and 7 T. (**H** and **I**) Differential resistance dV/dI as a function of $I_{DC}$ at out-of-plane magnetic field $B_{\perp}$ = 0 (H) and 15 mT (I), measured near where $T_c$ is highest (same positions in **g**) at $B_{||}$ = 0, 1, 3, 5, and 7 T. In (H), the resistance peaks shift to higher $I_{DC}$ at higher $B_{||}$. In (I), the resistance peaks disappeared for $B_{||} \leq 3$ T, indicating that superconductivity has already been fully suppressed at $B_{\perp}$ = 15 mT. In contrast, the resistance peaks are still visible at $B_{||}$ = 5 and 7 T, suggesting that SC5 is enhanced to survive from the same out-of-plane magnetic field.*

## Understanding the parent state of SC5

Figure 3B shows the longitudinal resistance $R_{xx}$ as a function of density $n$ and out-of-plane magnetic field $B_\perp$, corresponding to the red dashed line in Fig. 3a. By performing a fast Fourier transform (FFT) on $R_{xx}(1/B_\perp)$, as shown in Fig. 3C, we can utilize the fermiology analysis method(*14*, *30*, *39*) to extract the Fermi surface information near SC5. Note that we use only data in the range of $B_\perp$ = 0.02 T - 0.45 T, because a higher out-of-plane magnetic field will change the Stoner phases and thus affect the fermiology analysis. The pattern in Fig. 3C can be split into four regions, separated by the white vertical dashed lines, labeled as (a) ~ (d) from high density to low density. In each part, we identify the corresponding frequency peaks, highlighted by red dashed lines and named as $f_1$, $f_2$, $f_3$… in ascending order. In regions (a) and (b), three and two density-dependent peaks can be recognized, respectively. In regions (c) and (d), the FFT intensity is dominated by one single peak at $f_v = 1$ and $f_v = 1/6$, respectively.

The value of $f_v$ represents the relative size of Fermi pockets compared to the full size if there is only one Fermi surface(*39*). Therefore, in region (c), we can conclude there is only $1/f_1 = 1$ Fermi pocket, corresponding to a QM phase, which has been confirmed by previous studies(*24*, *26*). We can also deduce that in region (d), there are $1/f_1 = 6$ Fermi pockets of the same size, corresponding to a half-metal (HM) phase with trigonal warping(*60*). The Fermi surfaces in regions (a) and (b), however, are more complex. In region (b), though $f_1$ & $f_2$ both shift as $n$ changes, they always satisfy the relationship $6\times f_1 + 6\times f_2 \approx 1$. This can be understood as an unbalanced full-metal (FM) phase with trigonal warping: in each isospin flavor, there exist three tiny Fermi pockets, and Ising-type SOC leads to an imbalance between two groups of isospin flavors, so there are in total six large and six small Fermi pockets. Similarly, region (a) can be understood as an unbalanced three-quarter-metal (TQM) phase(*24*), since the FFT pattern satisfies $3\times f_1 + 3\times f_2 + f_3 \approx 1$. Figure 3D illustrates the Fermi pockets in all four regions, where blue and red color corresponds to the isospin flavors that are energetically more favorable or unfavorable by the Ising SOC. In Fig. 3A, we also mark their positions in the $n$-$D$ phase diagram.

Figure 3E shows a phase diagram obtained from a self-consistent Hartree-Fock calculation (see Supplementary Materials for details), which captures all the main phases in Fig. 3A and matches well with the fermiology analysis in Fig. 3C. Although in Fig. 3C we show that region (a) is in TQM phase, the whole valley-imbalanced (VI) phase(*55*) in Fig. 3A contains HM, TQM, and also FM phases. The fluctuations of $R_{xx}$ at zero magnetic field (see Fig. S1A & D) and the anomalous Hall effect observed in $R_{xy}$ measurement in Fig. 3F further confirm its valley-polarized nature. On the other hand, SC5 resides outside the tip of the QM phase, indicating that its parent states are the FM and HM phase. No hysteresis or anomalous Hall effect was observed in the $R_{xy}$ measurement at SC5 in these two phases (Fig. 3G and Fig. S5B & C), which again confirms that the parent state of SC5 is valley-unpolarized. The transition between the HM phase and the FM phase is continuous, as the less-populated red pockets in the FM phase can emerge smoothly. This explains why no phase boundaries can be observed in Fig. 3A and can be further verified by the additional fermiology analysis in Fig. S5.

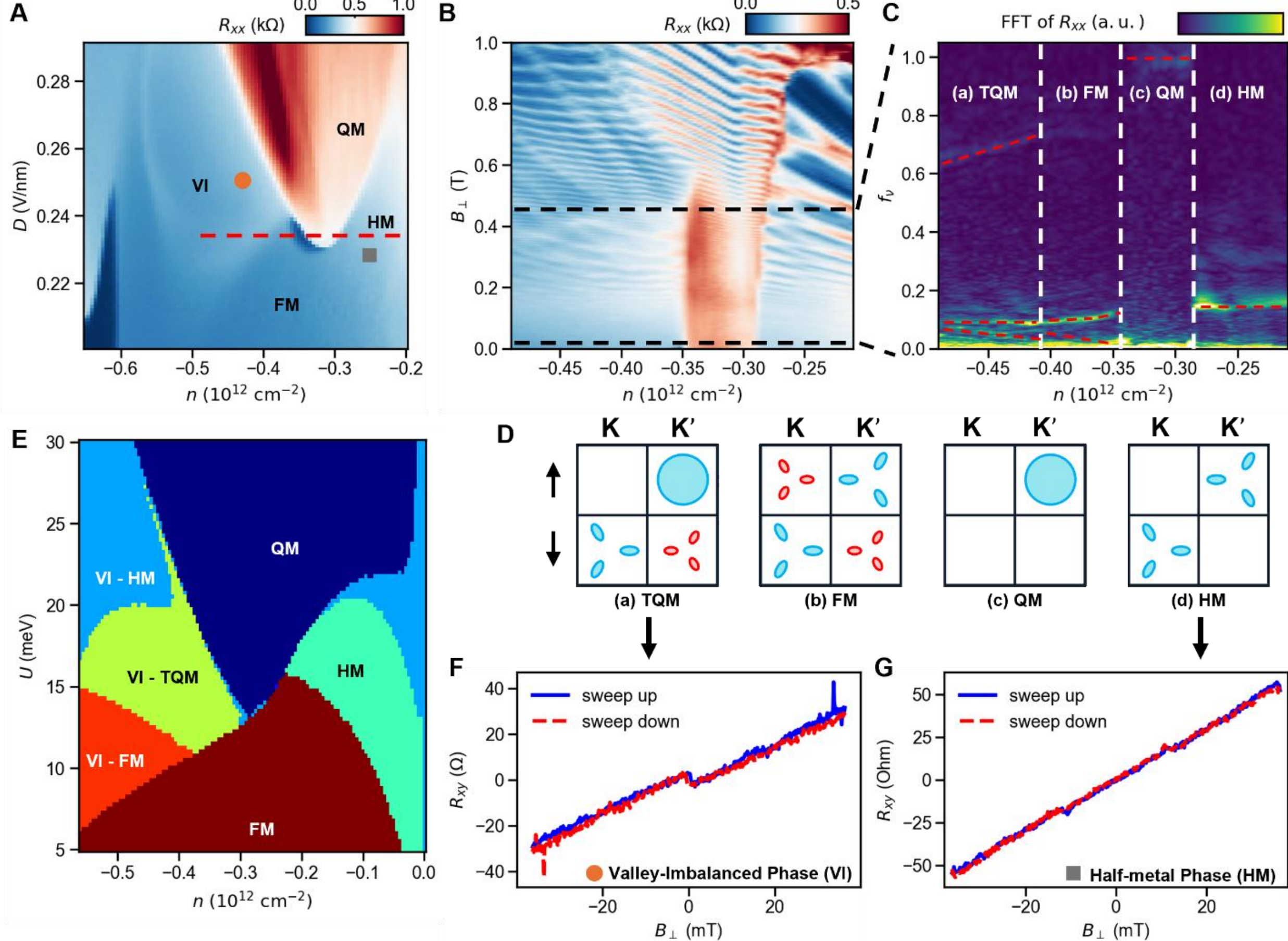

***Fig. 3: Fermiology analysis of SC5.***

*(**A**) Longitudinal resistance $R_{xx}$ as a function of n and D, covering SC5 and nearby isospin-symmetry-broken phases. VI stands for valley-imbalanced phase. (**B**) The dependence of $R_{xx}$ on $B_\perp$ and n, measured at $B_{||} = 0$, and $D = 0.235$ V/nm (corresponding to the red dashed line in (A)). Quantum oscillations of $R_{xx}$ are observed starting from $B_\perp = 0.16$ T. (**C**) Fourier transform of $R_{xx}(1/B_\perp)$ using the data from (B) within the range of $B_\perp = 0.02$ to $0.45$ T. The $f_v$ is defined as $f_{1/B}/\Phi_0 n$, where $\Phi_0 = h/e$ is the magnetic flux quantum. Three vertical white dashed lines mark the phase boundaries between the three-quarter-metal (TQM) phase, the full-metal (FM) phase, the quarter-metal (QM) phase, and the half-metal (HM) phase. Red dashed lines highlight all the frequency peaks used in Fermiology analysis for each region. In the TQM phase, three frequency peaks $f_1 \sim f_3$ (in an ascendant order, same for below) satisfy $3\times f_1+3\times f_2+f_3 \approx 1$. In the FM phase, $f_1$ & $f_2$ satisfy $6\times f_1+6\times f_2 \approx 1$. In the QM and HM phases, there is only one single peak at $f_1 \approx 1$ and $f_1 \approx 1/6$, respectively. (**D**) Schematic illustration of the Fermi surfaces for the four phases identified in (B). The blue / red color represents the isospin flavors that are more favorable / unfavorable in energy by Ising SOC effects. The parent state of SC5 is the FM or HM phase with trigonal warping. (**E**) Phase diagram near SC5 from Hartree-Fock calculations. All phases discovered in (C) were observed in the calculation. Note that the three phases on the high-density side are valley imbalanced (VI). (**F** and **G**) $R_{xy}$ measurements within the VI phase (F) and the HM phase (G), corresponding to the orange dot & grey square marker position in (A). Anomalous Hall effect was observed only in (F), illustrating its valley-imbalanced nature.*

**Discussions**

To quantify the extent to which the superconducting states break the Pauli limit, one usually defines the Pauli-limit violation ratio (PVR)(*24*, *43*, *44*, *46*, *61*) as the in-plane critical field at zero temperature divided by the Pauli limit calculated from the critical temperature at zero field: PVR = $B_{c\parallel}$(0 K) [T] / ($T_c$(0 T) [K] * 1.86). Simply applying this estimate to a fixed *n-D* position in Fig.1B, however, could mislead the analysis of how robust superconductivity is against magnetic field: the superconducting region could shift and expand in the *n-D* map with magnetic field (see Fig. 2D, for example), and a non-superconducting state at zero $B_\parallel$ can become a superconducting one at non-zero $B_\parallel$. In such cases the PVR becomes infinite and is no longer a meaningful metric. Therefore, we instead define $T_c$ to be the highest critical temperature among all *n-D* positions within SC5 at zero magnetic field, as it reflects the strength of SC5 regardless of its shift and expansion with $B_\parallel$. Specifically, $T_c$ is 68 mK in the first-round measurement (region 1), and 60 mK for the second-round measurements using another pair of $R_{xx}$ contacts (region 2). Using $B_\parallel$ = 8.835 T (at which SC5 can still be observed; see Fig. S6B) as $B_{c||}$, we obtain a record-high PVR value of 80 for region 2, higher than that in any other known superconductors. Figure 4a summarizes all other two-dimensional and three-dimensional superconductors(*7*, *8*, *11*, *12*, *14*, *22*, *25*, *44*, *46*, *52*, *53*, *61*–*65*) that also significantly break the Pauli limit to the best of our knowledge, and the two data points in our experiment (red star markers) lie much higher than all others. The true PVR could be higher than 80, since this is solely limited by the magnet available here.

Experimentally, we observe that SC5 extends into both the HM and FM phases. It is therefore reasonable to assume that in SC5 Cooper pairs are formed by electrons from their common HM parts, i.e., the six small blue Fermi pockets in Fig. 3D (b) and (d). These pockets feature a balanced valley occupation, as can be deduced from the absence of anomalous Hall signal in Fig. 3G. Based on the existence of Ising SOC and the self-consistent Hartree-Fock calculation (see Supplementary Materials for details), we can further conclude this valley-balanced HM also features spin-valley locking and a balanced spin occupation. Although intra-valley superconducting pairing has been experimentally observed in multilayer RG(*25*) accompanied by an anomalous Hall effect, here we consider that SC5 is still formed by the less exotic inter-valley pairing. We therefore determine that SC5 at zero magnetic field is an Ising-type superconductor.

As $B_\parallel$ is turned on, spin-canting towards the in-plane direction is expected to minimize the total free energy. We calculate the canting angle $\varphi$ in the normal state as a function of $B_\parallel$ in Fig. 4b by considering the competition among valley-interchange interactions, Ising SOC, and spin Zeeman energy. The strength of induced Ising SOC in our sample, $\lambda_I \approx 0.5$ meV, is determined from the beating patterns in the quantum oscillations (see Fig. S7) and from the previous analysis(*24*) of Ising-type SC4 in the same sample. At $B_\parallel$ = 8.835 T, $\varphi$ reaches ~ 71°, indicating that the parent state of SC5 is largely spin-polarized, although there is a small Ising component.

Given the enhancement of $T_c$ and critical out-of-plane magnetic field shown in Fig. 2, SC5 at high in-plane magnetic field clearly goes beyond the picture of an Ising superconductor with only spin-singlet pairing. This conclusion is supported by the fact that the critical $B_\parallel$ is far higher than that predicted by a spin-singlet-pairing-based Ising superconductivity ($B_{c,Ising} \sim \sqrt{B_{SOC} B_p} \approx 0.98$ T for region 2). Instead, as illustrated in Fig. 4C, we shall understand SC5 as a coherent superconducting superposition of spin singlets and spin triplets, which is allowed by the breaking of inversion symmetry in our device. We may explain the evolution of SC5 with $B_\parallel$ using the following simple phenomenological picture. At $B_\parallel$ = 0 T, the spin- and valley-balanced HM parent state only allows superconductivity in the spin singlet channel or in the spin-unpolarized triplet channel, while the latter is typically much weaker(*66*). As $B_{\parallel}$ is increased, the in-plane spin-polarized triplet channel becomes progressively possible through spin-canting. One may express the overall superconducting gap as follows(*51*, *67*): $\Delta = \sqrt{\Delta_s^2 + \Delta_t^2}$, where $\Delta_s$ and $\Delta_t$ are the spin-singlet and spin-triplet superconducting order parameters, respectively. Unlike superconductivities in two-dimensional TMDs, where $\Delta_t$ is much smaller than $\Delta_s$ or does not even exist(*66*, *67*), in SC5 $\Delta_t$ is playing a significant role, so that the superconductivity $T_c$ can be enhanced when the in-plane Zeeman field and the induced in-plane magnetization are increased(*51*). However, the detailed pairing mechanism in this in-plane spin-polarized triplet channel is unclear at this moment, calling for future investigation.

Finally, we would like to emphasize that SC5 is phenomenologically distinct from a few previously claimed spin-polarized superconductors in other graphene-based systems. For example, in RTG without any TMD proximity effect, SC2 was claimed to be a spin-polarized superconductor(*22*), but the PVR was limited to ~ 10 and no enhancement of $T_c$ by in-plane magnetic field was reported. In magic-angle twisted trilayer graphene(*61*), re-entrant superconductivity was observed, but that only occurs at some specific $n$ and $D$ positions. Some theoretical study(*59*) even suggests that no Pauli-limit violation exists due to its strongly-coupled nature. In Bernal bilayer graphene(*14*, *65*), superconductivity was induced by an in-plane magnetic field of 0.165 T, but it would ultimately be suppressed at around $B_{\parallel}$ = 3 T – equivalent to a PVR of 54. Nevertheless, the magnetic-field-enhanced behavior of SC5 with record-high PVR is unique. This opens a new avenue for creating and manipulating spin-polarized superconductivity in a clean, tunable, and crystalline system, while highlighting the need for further investigation into the underlying unconventional pairing mechanism.

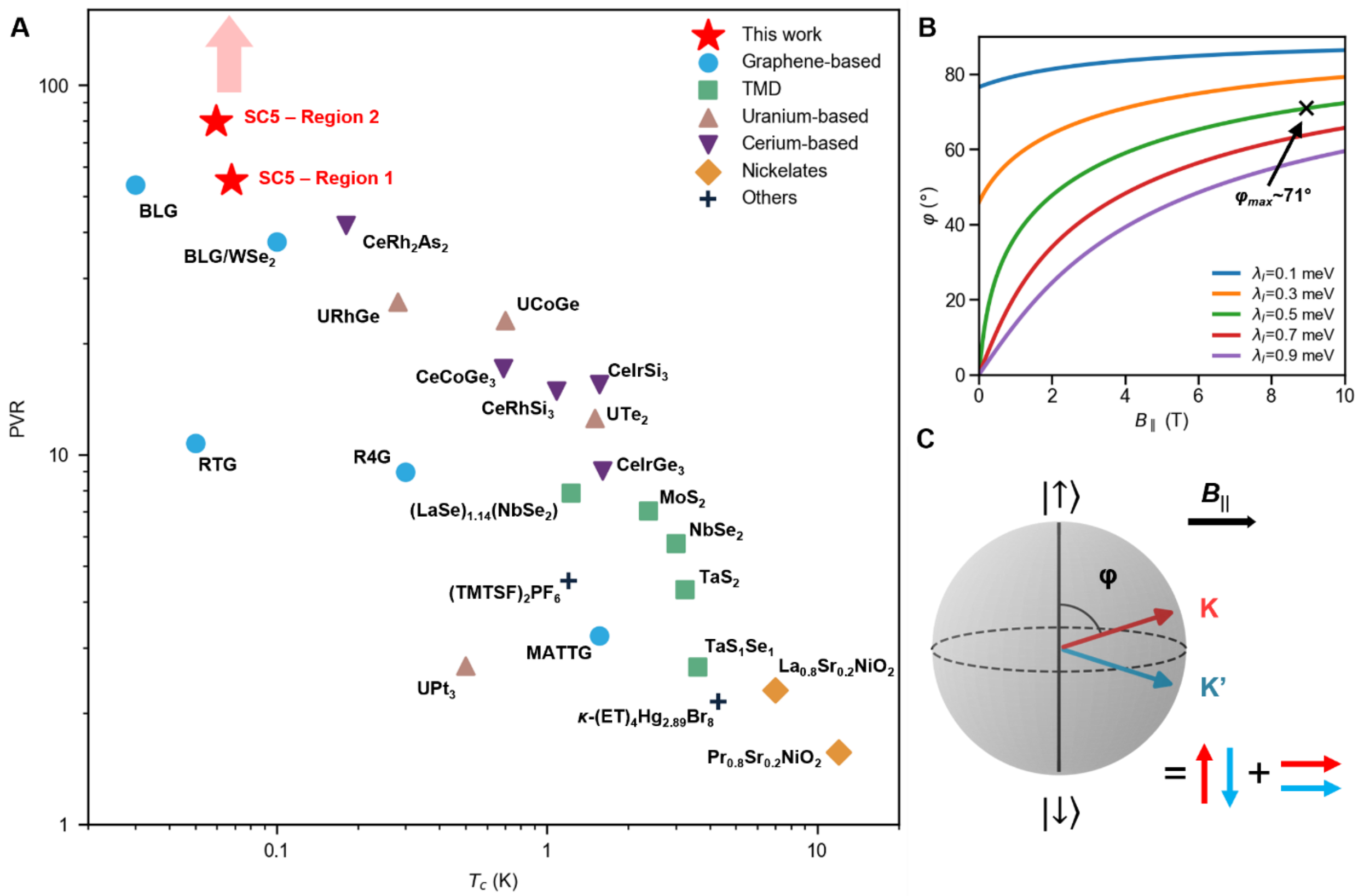

***Fig. 4: Record-high PVR of spin-polarized valley-unpolarized superconductivity.***
*(**A**) A summary of Pauli-limit violation ratio (PVR) vs. critical temperature $T_c$ at zero magnetic fields for different superconductors that have been reported to break the Pauli limit. Data is directly adapted or calculated from references(7, 8, 11, 12, 14, 22, 25, 44, 46, 52, 53, 61–65). For 2D superconductors, applied magnetic field is along the in-plane direction. For 3D superconductors, the direction of applied magnetic field can be along different crystal axes for crystals (or arbitrary for polycrystals), so we only present the highest PVR value of each material for simplicity. SC5 (marked as red stars) exhibits the highest Pauli-limit violation ratio (PVR) of 80, which is solely limited by our experimental instruments and could be even higher. (**B**) Calculated canting angle $\varphi$ as a function of in-plane magnetic field $B_{||}$ with different Ising SOC strength $\lambda_I$. For the maximum $B_{||}$ = 8.835 T, $\varphi_{max}$ reaches ~71° given $\lambda_I$ = 0.5 meV. (**C**) Illustration of the spin-canted nature of SC5 under finite in-plane magnetic field. The north and south poles of the Bloch sphere represent out-of-plane spin directions, while the red and blue arrows correspond to the spins in the K and K' valleys, respectively. Due to the existence of $B_{||}$, both spins tilt towards the in-plane direction to minimize the free energy. SC5 can then be understood as a coherent superconducting superposition of spin-unpolarized singlets and spin-polarized triplets, and it is dominated by the triplet component under high in-plane magnetic field.*

**Acknowledgement:**
We acknowledge helpful discussions with M. Meisel, L. Fu, E. Zeldov, E. Berg, L. Levitov, Z. Dong, Y. Zhang, T. Wang, Z. Dong, and J. Dong. **Fundings:** This work was mainly supported by ONR through Grant number N00014-25-1-2294. The device fabrication and transport measurements at MIT are supported by the Nano & Material Technology Development Program through the National Research Foundation of Korea (NRF) funded by the Ministry of Science and ICT (RS-2024-004447252), and the MIT Portugal Program. T.H. acknowledge a Mathworks Fellowship. The device fabrication for this work was carried out at the Harvard Center for Nanoscale Systems and MIT.nano. The National High Magnetic Field Laboratory (MagLab) is supported by the National Science Foundation through NSF/DMR-2128556 and by the State of Florida. Work in University of Basel was supported by the EU's H2020 Marie Skłodowska-Curie Actions (MSCA) Cofund Quantum Science and Technologies at the European Campus (QUSTEC) grant no. 847471, the Swiss National Science Foundation (grant no. 215757), the Georg H. Endress Foundation, the WSS Research Center for Molecular Quantum Systems (molQ) of the Werner Siemens Foundation and the UpQuantVal InterReg. K.W. and T.T. acknowledge support from the JSPS KAKENHI (Grant Numbers 21H05233 and 23H02052), the CREST (JPMJCR24A5), JST and World Premier International Research Center Initiative (WPI), MEXT, Japan. C.Y. and F.Z. (UT Dallas) were supported by the NSF under grants DMR-2414726, DMR-1945351, and DMR-2324033 and by the Welch Foundation under grant AT-2264-20250403. C.Y. and F.Z. acknowledge the Texas Advanced Computing Center (TACC) for providing supercomputing resources. **Author Contributions:** L. J., D. M. Z., and F. Z. supervised the project. J. Y., O. S. S., S. Y., H. W., A. C., T. H., Z. L., Z. H., J. S., M. X., C. S., M. Z., R. G., D. L., and M. L. performed the transport experiments. J. Y., S. Y., T. H., L. S., E. A., P. P. L., and Z. W. fabricated the samples. Z.L. performed preliminary measurements with assistance from R.G. at the MagLab HBT. C. Y. and F. Z. performed theoretical calculations. K. W. and T. T. grew hBN crystals. All authors discussed the results and wrote the paper. **Competing interests:** D. M. Z. and C. S. are co-founders of Basel Precision Instruments. The other authors declare no competing financial interests. **Data availability**: The data shown in the figures are available upon request.

**Supplementary Materials:**
Methods
Figs. S1 to S7
References (68-70)